# The impact of state capacity on the cross-country variations in COVID-19 vaccination rates

**Dragan Tevdovski**
Faculty of Economics, University Ss. Cyril and Methodius in Skopje, Skopje, North Macedonia

**Petar Jolakoski**
Association for Research and Analysis – ZMAI, Skopje, North Macedonia

**Viktor Stojkoski**
Faculty of Economics, University Ss. Cyril and Methodius in Skopje, Skopje, North Macedonia,
Association for Research and Analysis – ZMAI, Skopje, North Macedonia
Macedonian Academy of Sciences and Arts, Skopje, North Macedonia
vstojkoski@eccf.ukim.edu.mk

**Abstract**

*The initial period of vaccination shows strong heterogeneity between countries' vaccinations rollout, both in the terms of the start of the vaccination process and in the dynamics of the number of people that are vaccinated. A predominant thesis in the ongoing debate on the drivers of this observed heterogeneity is that a key determinant of the swift and extensive vaccine rollout is state capacity. Here, we utilize two measures that quantify different aspects of the state capacity: i) the external capacity (measured through the soft power and the economic power of the country) and ii) the internal capacity (measured via the country's government effectiveness) and investigate their relationship with the coronavirus vaccination outcome in the initial period (up to 30th January 2021). By using data on 189 countries and a two-step Heckman approach, we find that the economic power of the country and its soft power are robust determinants of whether a country has started with the vaccination process. In addition, the government effectiveness is a key factor that determines vaccine roll-out. Altogether, our findings are in line with the hypothesis that state capacity determines the observed heterogeneity between countries in the initial period of COVID-19 vaccines rollout.*

## 1. Introduction

The COVID-19 virus created the most severe global economic crisis since the Global Depression. The recent innovation of COVID-19 vaccines brings hope that the World will exit from the pandemic soon. As a means to provide an equitable access to vaccines for every country, the World Health Organization created the COVAX system. However, the system has so far failed in its goal, and the initial period of vaccination shows strong heterogeneity between countries' vaccinations rollouts. This is observed both in the terms of the start of the vaccination process and in the dynamics of the number of people that are vaccinated. Concretely, until end-January 2021, the vaccination started only in 54 countries in the World, while the COVID-19 vaccines are not accessible for most of the countries. Also, there are high discrepancies among the countries in speed of vaccination. Israel is leading with 46,7 administered COVID-19 doses per 100 people, followed by United Arab Emirates 27.1, United Kingdom 10.8 and United States 7.1, while the rest of the countries are below 5 administered COVID-19 doses per 100 people on January 30th, 2021.[1]

A predominant thesis in the ongoing debate on the drivers of this heterogeneity is that a key determinant of the swift and extensive vaccine rollout is *state capacity*. This variable is not strictly defined in the literature but is generally portrayed as a potential source of strength that can fundamentally shape the implementation and the final impact of policies (Cingolani et al. 2015). As such it has attracted renowned interest in studies of economic development over the past decade (for example, Besley and Persson 2010, Acemoglu et al. 2015, Cingolani 2018, Khemani 2019, Williams 2020).

In our work, we bridge the gap between the theoretical debate and empirical evidence by conducting a thorough econometric analysis on the impact of state capacity on the ability of a country to start and implement the vaccination process. For this purpose, we utilize country level data that covers the initial period of the vaccination process, up to January 30$^{th}$, 2021 and model the final outcome (the total number of administered vaccines per capita at the final date). Evidently, the data is censored i.e., most countries display 0 vaccination rate. To circumvent this problem, we assume that countries undergo a process of negotiation with vaccine providers and accept an offer only if it suits their economy. This allows us to describe the vaccination outcome as a two-step Heckman selection model. In the first step, we model the probability that a country has started with the vaccination process by using all available data. In the second step, we model the vaccination rate only for the countries that have started with the vaccination process.

We follow the literature and view the importance of the state capacity in two dimensions that determines COVID-19 vaccination success. The first is external capacity which describes the ability of the state to influence outcomes outside of its territory. In this case, it determines the likelihood that the country will secure reception of vaccines in a time period when vaccine supply is much lower than demand (i.e., a non-censored observation). The second is the internal state capacity. The internal capacity is relevant after the vaccines are received and it determines the speed of vaccination among a country's population.

The impact of the external state capacity on the supply of vaccines is based on the economic power and soft power of the country. As an approximation for economic power, we use the total GDP of the country. In the gravity models of international trade, this observable represents economic mass and is seen as a crude measure of economic power. An additional similar measure can be the export of the country, which is an indicator of the success of an economy in international trade. Nye (2004) defines soft power as the capacity for shaping the preferences of others through appeal and cooperation. While there is no widely accepted measure of the soft power and different rankings focused on different elements of interest, in this paper, we use the membership of the country in the top 30 countries according to the Soft Power 30 index (USC Center of Public Diplomacy, 2020). This is a formal index that ranks countries according to three primary dimensions of soft power: culture, political values and foreign policy, based on theoretical considerations of Nye (2011). For various applications of the index, we refer to Rutland and Kazantsev 2016, Lai 2019, and Kacziba 2020.

We measure the internal state capacity with the World Bank's indicator of government effectiveness. The indicator captures perception about the quality of policies formulation and implementation. As a consequence, it has been widely used as an indicator of the administrative capacity (for example, Lee and Whitford 2009, Hanson and Sigman 2013).

---

[1] Source: Ourworld in data, https://ourworldindata.org/covid-vaccinations.

After controlling for a large set of potential socio-economic and COVID-19 specific confounders, we find that the economic power of the country and its soft power are robust determinants of the vaccines receipts and start of the vaccination process in the initial period, which is characterized with very limited vaccines supply from producers. In addition, our results suggest that after the vaccines' supply is secured, the internal state capacity, measured by the government effectiveness, is a key factor that determines vaccine roll-out. Altogether, our findings are in line with the hypothesis that state capacity determines the observed heterogeneity between countries in the initial period of COVID-19 vaccines rollout.

The rest of the paper is structured as follows. In Section 2 we present an econometric model. In Section 3 we describe the data used for the testing of our hypotheses. In Section 4 we present our main findings. Section 5 sets out our conclusions.

## 2. Method

Countries undergo a process of negotiation with a vaccine provider to determine the quantity and price of vaccines. This process involves a great deal of geopolitics and, in the sense that most of the countries negotiate only with vaccine providers which are recognized by its geopolitical allies. This is evident by the fact that many countries from Eastern Europe in the observed period declined offers of Sinovac from China or Sputnic V from Russia, while failed to secure Western vaccines made by BioNTech-Pfizer, Moderna and AstraZeneca.[2] In addition, there might be skepticism in the quality of certain vaccines. For simplicity we assume that these aspects are translated in the price of the vaccine. The resulting outcome for country $i$ is the offer for the number of vaccinations per hundred population $v_i^*$, which is unobserved. If the offer is above a certain threshold that is acceptable for the price, the country will accept the offer. Moreover, the start of vaccination is conditioned with physical distribution of the vaccines in the observed period.[3] Evidently, there is a bias in the selection of the sample countries used for modeling the COVID-19 vaccination dynamics, as not everyone was able to efficiently negotiate with vaccine providers and to secure receipt of the vaccines.

A simple way to correct for this bias is to utilize a Heckman selection model (Heckman, 1974; Winship, 1992). The model represents a simple two-step approach. In the first step of the estimation procedure, the researcher models the sampling probability of each observation. In our case, this is the selection equation, where we quantify the probability that a certain country has started with the vaccination process. This is modeled as a probit regression of the form

$$Prob(V_i = 1 \,|X_{1i}, Z_{1i}) = \Phi(\beta_1 X_{1i} + \gamma Z_{1i}),$$

where $V_i$ is an indicator variable which takes value 1 if country $i$ has started with the vaccination process, and zero otherwise; $X_{1i}$ is a vector of potential socio-economic variables that determine the value of $V_i$, with $\beta_1$ being their marginal effects; and $Z_{1i}$ are control variables that are specific to the COVID-19 dynamics in the country, with $\gamma$ being their marginal effect. The second step models the observed vaccination rate by correcting the bias in the random error $u_i$ of the regression using the conditional expectation of the error estimated in the first step. Formally, the second step of the equation is specified as

$$E(v_i|\, X_{2i}, Z_{2i}, V_i = 1) = \beta_2 X_{2i} + \delta Z_{2i} + E(u_i|\beta_2 X_{2i} + \delta Z_{2i}).$$

---

[2] See the report accessed at https://ecfr.eu/article/the-geopolitics-of-covid-vaccines-in-europes-eastern-neighbourhood/
[3] In order to force pharmaceutical companies to fulfil COVID-19 contracts, European Union has introduced a temporary export regime for COVID-19 vaccines on January 30th, 2021.

In the equation, $X_{2i}$ is a vector of potential socio-economic variables that determine the vaccination rate of a country which has started the vaccination process, with $\beta_2$ being their marginal effects; and $Z_{2i}$ are control variables that are specific to the vaccination rate dynamics in the country.

## 3. Data

We obtained COVID-19 number of vaccinations per hundred population data from Our World in Data in Data – Coronavirus (COVID-19) Vaccinations database[4]. The explanatory variables were taken from various sources, as will be explained in the following. For each variable, unless otherwise specified, we take the last available data point as our observation, with the note that we do not take data that came out after January 30th 2021. This is because we want to emphasize the role of the country's capacity in being able to deliver COVID-19 vaccines in the initial period of the vaccination process.

In the first step of the model specification, we include two control variables that may govern the probability that a country has started with the vaccination process. They are the log of the registered COVID-19 cases up to the last date of observation and the log of the average daily government response index since the first registered case and up to the last date of observation. The first variable quantifies the magnitude of the health crisis in the country, whereas the second one is an estimate for the government behavior aimed at reducing the impact of the crisis. The data for the registered Cases were taken from the Our World in Data Coronavirus database[5]. The government response index was calculated using the methodology described in (Stojkoski et al., 2020) with input data from the Oxford Government Response Tracker[6]. In addition, the socio-economic variables that may drive the observation of whether a country has started with the vaccination process and are included in our model are: the log of the GDP of the country – measures the size of the economy and its economic power; the log of the share of exports in GDP – quantifies the presence of the country in the global trade market; the log of the total health expenditures as a percent of GDP – provides an proxy for the capacity of the health sector in the country, the log of the Military expenditures as a percent of GDP – determines the hard power of the country, and a dummy quantifying whether the country is included in the top 30 countries that are ranked in terms of soft power – obviously, measuring the soft power of the country.

In the second step, besides the control variables included in the first step we also include the number of days since the first vaccination in the country and dummies for three different types of vaccine providers present in the country i) Western block vaccines (Oxford/Az, Pfizer/BioNTech and Moderna), ii) Chinese vaccines (Sino-Wuhan) and iii) Russian vaccines (Sputnik). The first quantity is a measure for the duration of the vaccination process in the country, whereas the second one is an approximation for the heterogeneity in the vaccine choices of a country and indirectly it may be a quantity for the geopolitical orientation of the country. Moreover, in the second step we include each of the socio-economic variables except the soft power dummy, which is now substituted with our measure for the internal capacity of the country – the government effectiveness. Also, we include the log of the share of persons above 65 years of age in the total population of the country. We expect that countries with older populations also speed up with the vaccination process as they are the most susceptible group to the disease. Data sources, variable descriptions and their abbreviations are presented in more detail in Appendix, Table A1.

---

[4] Available at https://ourworldindata.org/covid-vaccinations.
[5] Available at https://ourworldindata.org/coronavirus.
[6] Available at https://www.bsg.ox.ac.uk/research/research-projects/coronavirus-government-response-tracker.

Column 1 of Table 1 gives the mean value for each variable included in the sample. Columns 2 and 3 of the same table, give the summary statistics divided by group of countries which respectively have started and have not yet started the vaccination process. In total, only 56 of the 189 countries included in the sample have started with the vaccination process. Interestingly, the countries which started with the vaccination process also reported a higher number of registered cases per capita and had a stricter government response. Moreover, they had higher GDP, larger exports as well as health and military expenditures as a percent of GDP, older population and had more efficient government. All of this suggests that there are significant discrepancies in the economic performance of the countries which started with the vaccination process and those that have not.

**Table 1. Descriptive statistics**

| Variable | All countries | Started vaccination | |
|---|---|---|---|
| | | No | Yes |
| Vaccinations p.h.p (log) | 0.55 | / | 0.55 |
| | (1.57) | / | (1.57) |
| COVID-19 cases p.m.p (log) | 8.47 | 7.75 | 10.21 |
| | (2.31) | (2.28) | (1.17) |
| Days (first vaccine - last vaccine) | 27.11 | / | 27.11 |
| | (10.24) | / | (10.24) |
| Government response (log) | 57.22 | 56.83 | 58 |
| | (11.56) | (12.76) | (8.70) |
| GDP (log) | 25.05 | 24.32 | 26.67 |
| | (2.24) | (2.04) | (1.76) |
| GDP per capita, PPP (log) | 9.41 | 8.92 | 10.49 |
| | (1.16) | (1.02) | (0.61) |
| Exports (% of GDP, log) | 3.57 | 3.45 | 3.84 |
| | (0.61) | (0.58) | (0.61) |
| Health expenditures (% of GDP, log) | 1.77 | 1.69 | 1.97 |
| | (0.43) | (0.44) | (0.36) |
| Military expenditures (% of GDP, log) | 0.39 | 0.33 | 0.51 |
| | (0.92) | (1.03) | (0.67) |
| Government effectiveness | -0.05 | -0.43 | 0.83 |
| | (0.99) | (0.84) | (0.74) |
| Population 65+ (log) | 0.55 | 0.41 | 0.86 |
| | (0.46) | (0.38) | (0.48) |
| **Number of countries** | 189 | 133 | 56 |

Note: Standard deviation in parentheses

## 4. Results

### 4.1 Descriptive Analysis

We begin the analysis with a graphical representation of the association between the economic power and the probability for the country to start with the vaccination process in the initial period (December 15$^{th}$ 2020 - January 30th 2021). In this aspect, Figure 1 compares the boxplot of the log of GDP of the countries that started with the vaccination with the boxplot of the log of GDP of the countries that did not start in the observed period. The group of countries that started the vaccination process not only have a higher median, minimum and maximum, but also its first quartile is almost equal to the third quartile of the group of countries that did not start the vaccination process.

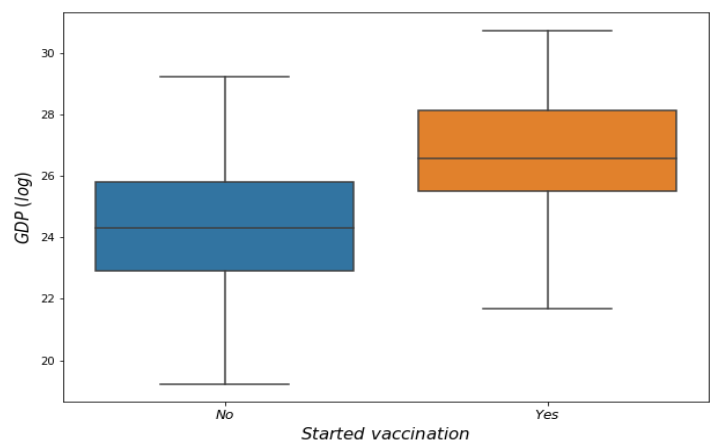

**Figure 1. Boxplots based on GDP.**

In more detail in Figure 2 we present with dots all 189 studied countries. Also, we plot the conditional probability density that a country started vaccination, given its GDP (in logs). The line shows that higher values of GDP lead to an increase in the probability that a country starts the vaccination process. Moreover, on this figure the countries that are in top 30 countries according to soft power are marked with orange. The vaccination started in 87% of the countries that are members of the Soft power 30. The four top soft power 30 countries where vaccinations did not start are characterized with low levels of daily COVID-19 cases (they are Australia, New Zealand, Japan and South Korea).

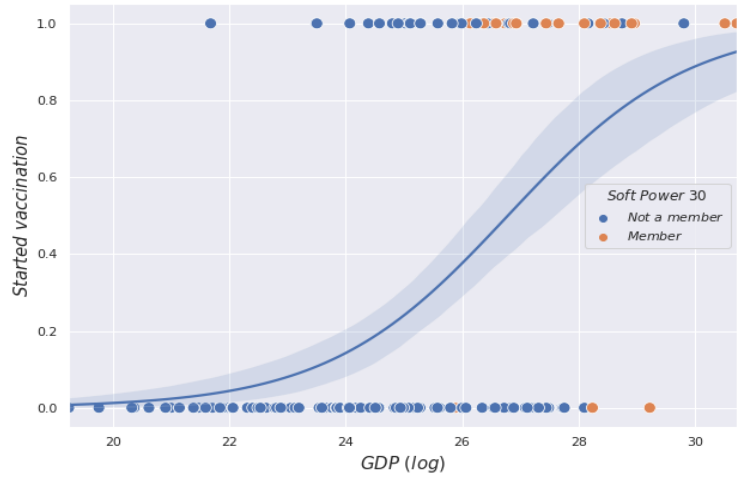

**Figure 2. Conditional probability density for a country to start the vaccination process given its GDP.** The filled background color indicates the 95% confidence interval.

In Figure 3, we present a scatterplot between government effectiveness and vaccination rate for all 56 countries that started with vaccination in the observed period. The plot uncovers a positive relationship between observed variables, implying that countries with higher government effectiveness achieved higher numbers of vaccinated persons, i.e., vaccination rate.

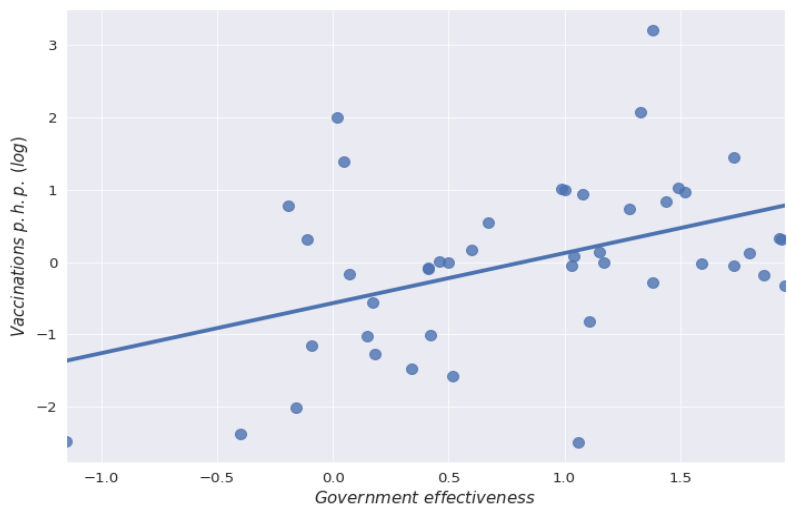

**Figure 3. Relationship between government effectiveness and vaccination rate**. The line shows the linear relationship between the variables (significant at 1%).

**4.2 Heckman selection model results**

Mirroring our theoretical model, the empirical analysis follows the described two-step process. In the first step, we quantify the probability for the start of the vaccination process in the observed period. The second step models the observed vaccination rate.

The results of the Heckman selection model are presented in Table 2. Columns (1-2) give the model estimates with only the control variables included in the model, Columns 3-12, represent results with four different specifications for the COVID-19 vaccination process. In each model, the dependent variable of the first step is marked with *started*, while in the second step with *lvac*. The estimated coefficients of the first model, as given in Columns 1 and 2 show that the registered COVID-19 cases are significant in both steps and that the number of days since the start of vaccination is significant in the second step, while the government response index is not significant. The positive coefficients imply that more registered COVID-19 cases in the country increases the probability country to start with the vaccinations in the observed period and also that more registered COVID-19 cases speed up the vaccination rate in the observed period. Also, as expected, more days since the start of vaccination lead to higher vaccination rate.

In the second model, given in Columns 3-4 of the same table, we add our measures of soft country capacity to the estimation. The results reveal that the country's membership in the top 30 soft power increases the probability that the country received vaccines and started with the vaccinations in the observed period. Also, the positive and significant government effectiveness coefficient implies that higher government effectiveness leads to higher vaccination rate in the country.

The third model (Columns 5-6) includes additional variables in both steps in order to check whether the capacity of the health sector or military power of the country influences the vaccination process. The estimated coefficients show that the capacity of the health sector in the country (the log of the total health expenditures as a percent of GDP) and the hard power of the country (the log of the military expenditures as a percent of GDP) are not significant determinants of the probability country to started with vaccination in observed period, while the soft power is still significant and positive. Similarly, these additionally added variables do not display a significant effect on the vaccination rate in the country.

The fourth model (Columns 7-8) is augmented with economic and demographic variables. In the first step, it additionally includes the log of the GDP and the log of the share of exports in GDP, while in the second step it additionally includes log of the GDP and share of the population above 65-years as the riskiest population group. The estimated coefficients in the first step show that both economic variables are significant and positive, implying that the size of the economy and the presence of the country in the international trade matter for both the probability that the country started with the vaccination process and the speed with which it distributes. Also, the soft power remains positive and significant. Additionally, the estimated coefficients in the second step show that the size of the economy and the share of the population above 65 years age are not significant, as well as previously added variables, while the days since the start of vaccination and government effectiveness remains significant and positive.

The estimated coefficients in each of the Heckman selection models back our theoretical claims. In every case, the economic power increases the probability that a country will secure the receipt of vaccines in the initial period of the start of vaccination. The bigger size of the economy represents a bigger market for the vaccine supplier and gives the country a better negotiation position. Also, stronger relative export capacity of the economy represents more importance in international trade and improves the prospects of the country to receive vaccines. Moreover, economic power is not a sole determinant of vaccine receipt, but also the soft power matters. We argue that the reason for this is not only limited supply of the vaccines in the period from December 15th 2020 to January 30th 2021, but also the fact that the competition for vaccines is not between private subjects, but between governments, which use their diplomatic and other apparatus. In addition, it is important to note that more registered COVID-19 cases mean more pressure on governments to find a way to secure the vaccines receipt in the initial period. Similarly, once the vaccines are secured, the speed of the vaccination depends on the government effectiveness to organize and execute the process of vaccinations, as well as the time of the realization of the process.

In the last model, we substitute the total GDP of the country with the GDP per capita and use it to check the significance of the predictors. In this case, the soft power variable is again a positive and significant predictor of the start of the vaccination, together with the number of registered COVID-19 cases and GDP per capita, which is an alternative measure of the economic power of the country, while the other variables remain not significant as in the previous models. In the second step, we observe that a significant predictor of the vaccination rate is only the number of the days since the first start of vaccination, whereas the newly included GDP per capita and the only variable that was previously significant, the government effectiveness, are not able to explain the COVID-19 vaccination dynamics together. However, the lost significance of the government effectiveness can be explained with high correlation between GDP per capita and government effectiveness (0.83). In Figure A1 in the appendix, we present a correlation matrix between all included variables in the analysis.

Table 2: Estimated Heckman selection models for the vaccination process.

| VARIABLE | Model 1 | | Model 2 | | Model 3 | | Model 4 | | Model 5 | |
|---|---|---|---|---|---|---|---|---|---|---|
| | (1) Vaccinations p.h.p. (log) | (2) Started vaccination | (3) Vaccinations p.h.p. (log) | (4) Started vaccination | (5) Vaccinations p.h.p. (log) | (6) Started vaccination | (7) Vaccinations p.h.p. (log) | (8) Started vaccination | (9) Vaccinations p.h.p. (log) | (10) Started vaccination |
| cases (log) | 0.361** | 0.525*** | 0.297** | 0.504*** | 0.315*** | 0.567*** | 0.304** | 0.469*** | 0.254** | 0.387** |
| | (0.165) | (0.130) | (0.123) | (0.096) | (0.121) | (0.145) | (0.146) | (0.120) | (0.128) | (0.150) |
| days | 0.087*** | | 0.069*** | | 0.085*** | | 0.071*** | | 0.068*** | |
| | (0.018) | | (0.015) | | (0.020) | | (0.018) | | (0.019) | |
| gov_response (log) | -0.900 | -0.452 | | | -0.889 | -0.205 | -0.356 | 0.086 | -0.502 | 0.021 |
| | (1.276) | (0.517) | | | (1.232) | (0.716) | (1.287) | (0.818) | (1.334) | (0.750) |
| GDP (log) | | | | | | | -0.178 | 0.400*** | | |
| | | | | | | | (0.199) | (0.127) | | |
| gdp_ppp_pc (log) | | | | | | | | | 0.574 | 0.737*** |
| | | | | | | | | | (0.697) | (0.281) |
| exports (log) | | | | | | 0.320 | | 1.090*** | | -0.022 |
| | | | | | | (0.309) | | (0.378) | | (0.344) |
| health_exp (log) | | | | | -0.328 | -0.521 | 0.037 | 0.044 | -0.439 | -0.323 |
| | | | | | (0.560) | (0.515) | (0.819) | (0.555) | (0.543) | (0.599) |
| military_exp (log) | | | | | 0.239 | 0.180 | 0.281 | -0.005 | 0.245 | 0.124 |
| | | | | | (0.267) | (0.193) | (0.215) | (0.202) | (0.233) | (0.177) |
| gov_eff | | | 0.718*** | | | | 0.782*** | | 0.335 | |
| | | | (0.217) | | | | (0.255) | | (0.525) | |
| pop_65 (log) | | | | | | | -0.221 | | -0.003 | |
| | | | | | | | (0.427) | | (0.406) | |
| Soft Power 30 | | | | 2.129*** | | 2.337*** | | 1.224** | | 1.286*** |
| | | | | (0.352) | | (0.435) | | (0.487) | | (0.368) |
| Constant | -2.022 | -3.408* | -5.120*** | -5.538*** | -0.878 | -5.519 | 1.746 | -19.707*** | -6.979 | -10.902** |
| | (5.632) | (1.872) | (1.213) | (0.935) | (5.597) | (3.423) | (10.182) | (6.337) | (6.517) | (4.322) |
| Observations | 165 | 165 | 187 | 187 | 151 | 151 | 148 | 148 | 148 | 148 |

Robust standard errors in parentheses. Every model includes dummies for the three types of vaccines i) Western block, ii) Chinese and iii) Russian.
*** p<0.01, ** p<0.05, * p<0.1

In Tables 3 and 4 we test the robustness of our results by performing two series of outliers check. In particular, in Table 3 we remove the countries that are below the 5th and above the 95th percentile in terms of the magnitude of government effectiveness and GDP, whereas in Table 4 we remove the countries that are above the 95th percentile in terms of the observed vaccination rate up until January 30th. In both cases, we repeat the estimation of the first four models discussed above, and each time our measures of soft power and external capacity remain significant predictors with a positive magnitude of, respectively, the probability that a country will start with the vaccination process and the vaccination rate in the initial period. Hence, our results are robust against the presence of outliers.

**Table 3: Estimated Heckman selection models (gov. effectiveness and GDP outliers excluded).**

| VARIABLES | Model 1 | | Model 2 | | Model 3 | | Model 4 | |
|---|---|---|---|---|---|---|---|---|
| | (1) Vaccinations p.h.p. (log) | (2) Started vaccination | (3) Vaccinations p.h.p. (log) | (4) Started vaccination | (5) Vaccinations p.h.p. (log) | (6) Started vaccination | (7) Vaccinations p.h.p. (log) | (8) Started vaccination |
| cases (log) | 0.518* | 0.821*** | 0.113 | 0.692*** | 0.711* | 0.600*** | 0.228 | 0.504*** |
| | (0.309) | (0.169) | (0.316) | (0.143) | (0.383) | (0.160) | (0.487) | (0.136) |
| days | 0.083*** | | 0.065*** | | 0.070** | | 0.063** | |
| | (0.024) | | (0.018) | | (0.035) | | (0.028) | |
| gov_response (log) | -1.642 | -0.618 | | | -1.537 | -0.187 | -0.824 | 0.120 |
| | (1.788) | (0.704) | | | (1.656) | (0.808) | (1.749) | (0.915) |
| GDP (log) | | | | | | | -0.243 | 0.313** |
| | | | | | | | (0.237) | (0.140) |
| exports (log) | | | | | | 0.667 | | 1.284** |
| | | | | | | (0.460) | | (0.561) |
| health_exp (log) | | | | | -1.575* | 0.214 | -1.134 | 0.560 |
| | | | | | (0.957) | (0.594) | (1.413) | (0.680) |
| military_exp (log) | | | | | 0.040 | 0.333 | 0.040 | 0.159 |
| | | | | | (0.347) | (0.240) | (0.258) | (0.231) |
| gov_eff | | | 1.405*** | | | | 1.447*** | |
| | | | (0.454) | | | | (0.535) | |
| pop_65 (log) | | | | | | | -0.166 | |
| | | | | | | | (0.473) | |
| Soft Power 30 | | | | 1.735*** | | 1.599*** | | 0.879* |
| | | | | (0.307) | | (0.355) | | (0.525) |
| Constant | -1.209 | -5.795* | -3.442 | -7.327*** | 0.834 | -8.669* | 8.852 | -19.759** |
| | (9.037) | (2.998) | (3.466) | (1.443) | (9.224) | (4.832) | (13.238) | (8.104) |
| Observations | 131 | 131 | 142 | 142 | 123 | 123 | 123 | 123 |

Robust standard errors in parentheses. Every model includes dummies for the three types of vaccines i) Western block, ii) Chinese and iii) Russian.
*** p<0.01, ** p<0.05, * p<0.1

**Table 4: Estimated Heckman selection models (vaccination rate outliers excluded).**

| VARIABLES | Model 1 | | Model 2 | | Model 3 | | Model 4 | |
|---|---|---|---|---|---|---|---|---|
| | (1) Vaccinations p.h.p. (log) | (2) Started vaccination | (3) Vaccinations p.h.p. (log) | (4) Started vaccination | (5) Vaccinations p.h.p. (log) | (6) Started vaccination | (7) Vaccinations p.h.p. (log) | (8) Started vaccination |
| cases (log) | 0.425*** | 0.516*** | 0.290*** | 0.496*** | 0.284** | 0.556*** | 0.331*** | 0.488*** |
| | (0.145) | (0.130) | (0.105) | (0.099) | (0.115) | (0.142) | (0.092) | (0.125) |
| days | 0.088*** | | 0.070*** | | 0.079*** | | 0.064*** | |
| | (0.016) | | (0.013) | | (0.018) | | (0.017) | |
| gov_response (log) | -0.490 | -0.403 | | | -0.250 | -0.084 | 1.091 | 0.242 |
| | (1.081) | (0.511) | | | (1.017) | (0.732) | (0.982) | (0.881) |
| GDP (log) | | | | | | | 0.042 | 0.498*** |
| | | | | | | | (0.101) | (0.118) |
| exports (log) | | | | | | 0.276 | | 1.163** |
| | | | | | | (0.298) | | (0.454) |
| health_exp (log) | | | | | 0.195 | -0.497 | -0.583 | 0.107 |
| | | | | | (0.473) | (0.518) | (0.591) | (0.605) |
| military_exp (log) | | | | | 0.076 | 0.158 | 0.265 | -0.020 |
| | | | | | (0.199) | (0.202) | (0.161) | (0.215) |
| gov_eff | | | 0.573*** | | | | 0.519** | |
| | | | (0.182) | | | | (0.217) | |
| pop_65 (log) | | | | | | | 0.633** | |
| | | | | | | | (0.315) | |
| Soft Power 30 | | | | 2.197*** | | 2.407*** | | 1.191** |
| | | | | (0.352) | | (0.416) | | (0.488) |
| Constant | -4.431 | -3.545* | -4.660*** | -5.526*** | -3.570 | -5.856 | -11.672** | -23.542*** |
| | (4.492) | (1.870) | (1.063) | (0.966) | (4.507) | (3.602) | (5.432) | (6.985) |
| Observations | 162 | 162 | 184 | 184 | 148 | 148 | 145 | 145 |

Robust standard errors in parentheses. Every model includes dummies for the three types of vaccines i) Western block, ii) Chinese and iii) Russian.
*** p<0.01, ** p<0.05, * p<0.1

## 5. Conclusion

We investigated the potential impact of state capacity on the ability of a country to start and implement the vaccination process in the initial COVID-19 vaccination period from December 15[th] 2020 to January 30[th] 2021. This was done by considering variables that can determine the likelihood that the country will secure reception of vaccines in a time period when vaccine supply is much lower than demand, as well as the variables that possibly affect the speed of the vaccination once the vaccines are received. By utilizing a Heckman two-step selection approach, in the first step, we quantified the probability for a certain country to start with the vaccination process in the observed period, based on the sample of 189 countries. In the second step, we modeled the observed vaccination rate in the countries that started with vaccination. The results of the model showed that economic power increases the probability that a country will secure the receipt of vaccines in the initial period of the start of vaccination. However, this is not the sole determinant, and the soft power of a country also matters. Moreover, once the vaccines are secured, we found that the speed of the vaccination is related with the government's effectiveness to organize and execute the process of vaccinations, as well as the time of the realization of the process. On the other hand, the country's policy response, total health expenditures, military expenditures, and proportion of the population above 65 years age are not significant. We also performed robustness checks, which were in line with the baseline results.

These results put a new light on the already known wisdom that the role of the state is crucial in a severe crisis, such as world wars, systemic economic crisis and pandemics. This particular time, during the COVID-19 pandemic, the vaccine rollout is a game changer as it reduces the health risks and directly impacts the speed and extent of economic recovery. Therefore, all countries around the world entered in a strong competition for vaccine receipt and implementation of the vaccination process in the initial period, which was characterized also by the non-functioning of the COVAX system, the global response system which was created with the aim to secure fair distribution of the vaccines in the countries around the world. In this aspect, the results of this paper are a stark reminder for the need for transparent and fair global response regarding fair and equitable availability of vaccines to every country.

Appendix

Table A1: Data sources and description of variables.

| Variable | Code | Definition | Data Source | Note |
|---|---|---|---|---|
| COVID-19 cases p.m.p | cases | Confirmed COVID-19 cases per million population | Our World in Data, Coronavirus | Measured in logs |
| Started vaccination | / | Indication whether the country has started with the vaccination procedure. | Our World in Data, Coronavirus | Binary |
| Vaccinations p.h.p | / | COVID-19 vaccination does administered per hundred people | Our World in Data, Coronavirus | Measured in logs |
| Government response | gov_response | Oxford's daily government response index measures the variation in daily government responses to COVID-19. | Oxford COVID-19 Government Response Tracker, Blavatnik School of Government | Measured in logs |
| Days | days | Number of days between the first vaccine and last available data for vaccinations p.h.p. | Our World in Data, Coronavirus | |
| GDP | gdp | GDP at purchaser's prices is the sum of gross value added by all resident producers in the economy plus any product taxes and minus any subsidies not included in the value of the products. | WDI, World Bank | Measured in logs |
| GDP per capita, PPP | gdp_pc_ppp | This indicator provides per capita values for gross domestic product (GDP) expressed in current international dollars converted by purchasing power parity (PPP) conversion factor. | WDI, World Bank | Measured in logs |
| Exports (% of GDP) | exports | Exports of goods and services represent the value of all goods and other market services provided to the rest of the world. | WDI, World Bank | Measured in logs |
| Health expenditures (% of GDP) | health_exp | Level of current health expenditure expressed as a percentage of GDP. | WDI, World Bank | Measured in logs |
| Military expenditures (% of GDP) | military_exp | Military expenditures data from SIPRI are derived from the NATO definition, which includes all current and capital expenditures on the armed forces. | WDI, World Bank | Measured in logs |
| Government effectiveness | gov_eff | Captures perceptions of the quality of public services, civil service and the degree of its independence from political pressures, quality of policy formulation and implementation, and credibility of the government | WGI, World Bank | |
| Population (65+, % of total) | pop_65 | Population ages 65 and above as a percentage of the total population. | WDI, World Bank | Measured in logs |
| Soft Power 30 | / | Soft power is the ability to attract and co-opt, rather than coerce (in contrast to hard power). | Soft Power 30 | Binary |

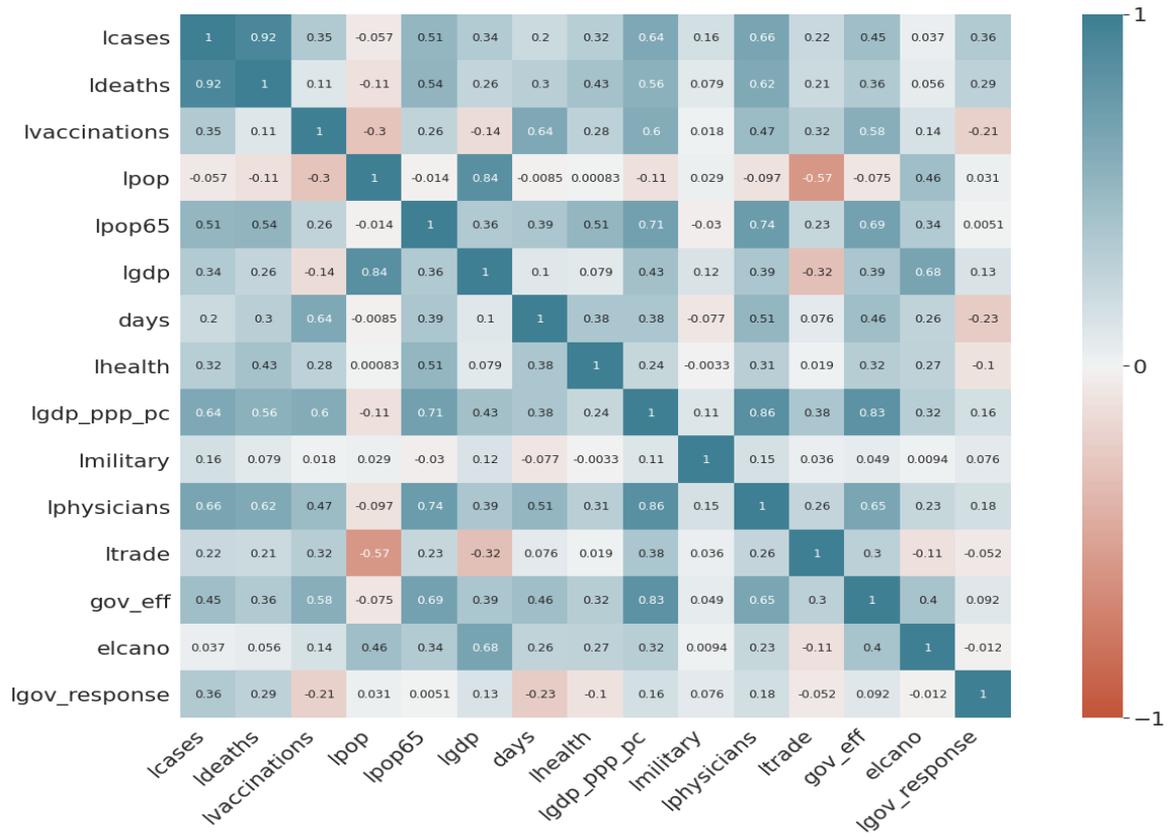

**Figure A1. Correlation matrix between the studied variables.**